\documentstyle[prd,preprint,tighten,aps,floats,eqsecnum,%
amssymb,amsfonts,newlfont,graphicx]{revtex}
\begin{document}
\draft
\preprint{
\begin{tabular}{r}
   DFTT 11/00
\\ arXiv:hep-ex/0003001
\end{tabular}
}
\title{What is the Usefulness of Frequentist Confidence Intervals?}
\author{Carlo Giunti}
%\email{Carlo Giunti <giunti@to.infn.it>}
\address{INFN, Sez. di Torino, and Dip. di Fisica Teorica,
Univ. di Torino, I--10125 Torino, Italy}
\date{March 1, 2000}
\maketitle
\begin{abstract}
The following questions are discussed:
``Why confidence intervals are a hot topic?'';
``Are confidence intervals objective?'';
``What is the usefulness of coverage?'';
``How to obtain useful information from experiment?'';
``The confidence level
must be chosen independently from the knowledge of the data?''.
\end{abstract}

The problem of getting meaningful information
from the statistical analysis of experimental data
in high energy physics
has attracted recently great attention,
reaching a (local) maximum
at the Workshop on ``Confidence Limits''
held at CERN in January
\cite{clw}.
Having participated to that Workshop
and having read some of the related material available at the Workshop
Web page
\cite{clw},
I think that there is a certain confusion
on the usefulness of Frequentist confidence intervals
and some clarifications are necessary.

In the following I will consider and answer
some crucial questions in the framework
of the Frequentist theory of statistical inference.
I will assume that the reader is familiar with the theory
and its problems
(if not, see \cite{clw,Neyman-37,Eadie-71,Cousins-95,PDG-96,%
Feldman-Cousins-98,PDG-98,Giunti-bo-99,Giunti-back-99,Giunti-98,%
Ciampolillo-98,Roe-Woodroofe-99,Mandelkern-Schultz-99,Punzi-99,%
Cousins-00,Giunti-Laveder-00,%
Cousins-clw,Giunti-clw,Zech-clw,Panel-clw}).

\begin{enumerate}

\item
\textbf{Why confidence intervals are a hot topic?}
\label{hot}

The current debate on the methods
of statistical analysis of experimental data
follows mainly from the proposal at the end of 1997
of the the Unified Approach by Feldman and Cousins
\cite{Feldman-Cousins-98}
and its immediate adoption as the recommended Frequentist method
in the 1998 edition of the
\textit{Review of Particle Physics} (RPP)
of the
Particle Data Group (PDG)
\cite{PDG-98}.
Although it may be true that
``there is no PDG method''
\cite{Groom-Panel-clw},
it is a matter of fact that RPP is a guide for the Physics community.
Most physicists have faith in what is written in RPP,
especially regarding the fields in which they are not experts
(unfortunately most human beings can achieve expertise in one or a few fields
and it is naturally correct to believe to experts in the other fields
if nothing that they say is obviously wrong).
Therefore,
the authors of RPP have a responsibility for what they write.

The immediate adoption of the Unified Approach by the PDG
has been considered by many rather premature,
taking into account that it happened before
testing the performances of the Unified Approach
in real experiments.
These concerns received dramatic confirmations from
the unphysical results obtained in two
of the first applications
\cite{KARMEN-nu98,Baudis-99}.
Several papers
\cite{Giunti-bo-99,Ciampolillo-98,Roe-Woodroofe-99,%
Mandelkern-Schultz-99,Punzi-99}
have followed the one by Feldman and Cousins,
proposing alternative Frequentist methods.
Hence,
at present there are several Frequentist approaches available
and each analyzer of experimental data
must choose one among them independently of the knowledge of the data,
in order to preserve the property of coverage
\cite{Feldman-Cousins-98}.
One of the main issues
in the present debate on confidence intervals
is the study of the properties of the different Frequentist methods
in order to allow a meaningful choice of the method to be used
in a practical application.

Another problem with the 1998 edition of RPP
is that
the emphasis on the Unified Approach
is likely correlated with the disappearance of
the useful description of the Bayesian approach
present in the 1996 edition of RPP
\cite{PDG-96}.
It seems hard to argue that this is not a biased choice.

\item
\textbf{Are confidence intervals objective?}
\label{objective}

It is well known that credibility intervals obtained in the framework
of the Bayesian theory
(see, for example, \cite{DAgostini-YR3-99})
are subjective because of the necessity to
have a prior probability distribution function
for the quantity under measurement.

In the Frequentist--Bayesian debate
some experts biased towards the Frequentist approach say that
they want to know what was measured in the experiment,
without the subjective Bayesian prior of the experimenter.
But also in the Frequentist approach
the experimenter must choose a method
to construct the confidence belt
and the result that he will obtain depends on this choice
(in Ref.~\cite{Giunti-Laveder-00}
it has been shown that Frequentist confidence intervals
are objective only from a statistical point of view).
Thus,
it is clear that Frequentist confidence intervals
(as Bayesian credibility intervals)
do not describe what is measured independently from subjective choices!

Actually,
in the framework of the Frequentist theory
there is no way to get a result without the subjective
choice of the method
to construct the confidence belt
(approximate methods as maximum likelihood
also need subjective choices).
On the other hand,
working in the framework of the Bayesian theory,
one can present the likelihood function
as the result of the experiment
(or its normalized version called
``relative belief updating ratio''
\cite{R})
and everyone can obtain a credibility interval using her/his prior.
Moreover,
the Bayesian prior takes into account in a proper way
the subjective believe that may come from
a solid experience in the field,
whereas the choice of a specific Frequentist method
seems much more arbitrary.

Since the Frequentist confidence intervals
are subjective as the Bayesian credibility intervals,
but Bayesian theory takes into account
subjective belief based on experience
in a proper way,
I think that,
\emph{contrary to what is usually believed,
a choice of the method based on subjectivity
favors the Bayesian approach}.
A choice of the Frequentist approach
is reasonable only if based on the main property
of Frequentist confidence intervals,
\emph{coverage},
that I will discuss in the following items.

\item
\textbf{What is the usefulness of coverage?}
\label{usefulness}

Some experts say that coverage implies that
in order to be right, for example, 90\% of the times,
10\% of the times you must get a wrong result and you must give it,
even if you know (or have a strong suspicion) that it is wrong
(for example, an empty confidence interval).
If I tell this to any pedestrian,
he will think that I am nuts:
if I know that the result is wrong why should I give it?
It is not only useless, it may also confuse other people.
So,
I think that among reasonable people we can agree that
\emph{wrong results are useless and should not be given}.

I think that coverage is useful because one knows the probability,
given by the confidence level,
that the confidence interval
covers the true value of the quantity under measurement.
Each confidence interval obtained in an experiment
has this property,
independently from the results and even existence of other experiments.
Therefore,
there is no need to give wrong confidence intervals!

In principle,
if one could make many experiments to measure a physical quantity $\mu$,
each experiment producing a confidence interval with a chosen confidence level,
one could collect all the confidence intervals
(including those that are known to be wrong),
producing a set of intervals
that cover the true value of $\mu$
with a probability given by the confidence level.
But in practice,
at least in high energy physics,
there are only a few experiments
(sometimes one or two)
that measure each physical quantity.
Therefore,
the set of confidence intervals is too small to be of any usefulness.
Instead,
one is interested to get useful information from each experiment.

\item
\textbf{How to obtain useful information from experiment?}
\label{information}

I think that a procedure that allows to get always useful information
from an experiments is the following:

\begin{enumerate}

\item
Choose the Frequentist method with the desired properties
independently of the knowledge of the data
(see \cite{Giunti-Laveder-00}).

\item
If the data do not indicate any unlikely statistical fluctuation
and
the confidence interval obtained with the chosen Frequentist method
looks fine,
the confidence interval
can be given and one knows that it covers the true value of the quantity
under measurement with a probability given by the confidence level.

\item
If it is clear that the data indicate an unlikely statistical fluctuation
(as less events than the expected background
measured in a Poisson process with known background)
and
the confidence interval obtained with the chosen Frequentist method
is suspected to be wrong
(for example, too small or even vanishing),
the confidence interval should not be given.

Feldman and Cousins
\cite{Feldman-Cousins-98}
proposed that in such cases the experimenter
should give also what they called
``sensitivity'',
but should be called more appropriately
\emph{``exclusion potential''}
\cite{Giunti-Laveder-00},
because it is calculated assuming that the quantity
under measurement is zero.
However,
since the exclusion potential
cannot be combined with the confidence interval
that has been obtained in the experiment,
it is not clear what is the usefulness of giving
two quantities instead of one,
except as a warning that the confidence interval
is likely to be wrong.
But in this case it is better not to give it!
Two quantities produce only confusion if one of them
tells that the other is not reliable.
Thus,
the solution proposed by Feldman and Cousins
(recommended also in the 1998 edition of RPP \cite{PDG-98})
to give two quantities instead of one is just the opposite
of what it is reasonable to do:
give nothing!
(Here I am discussing only Frequentist quantities.
One can always give Bayesian quantities,
as discussed in the next item.)

\item
In any case
the experimenters should
analyze their data using the Bayesian theory,
that allows always to obtain meaningful results.
The experimenters can give the
likelihood function or the relative belief updating ratio
\cite{R},
that represent the objective result of the experiment,
and
can give also a credibility interval
obtained with their prior
based on experience and knowledge of the experiment.

\end{enumerate}

Following this procedure,
experiments will always produce a result in the framework of the
Bayesian theory
and will produce also a Frequentist result only
if it is a reliable one.

As an illustration,
let us consider the well known case of the KARMEN experiment
on the search for short-baseline
$\bar\nu_\mu\to\bar\nu_e$ oscillations
\cite{KARMEN-nu98}.
In the middle of 1998
the KARMEN collaboration
reported the observation of
zero events in a Poisson process with a known background of
$2.88 \pm 0.13$
events
\cite{KARMEN-nu98}.
Using the Unified Approach
they obtained an exclusion curve
in the space of the neutrino mixing parameters
that seemed to exclude almost all the region allowed by the
positive results of the LSND $\bar\nu_\mu\to\bar\nu_e$ experiment
\cite{LSND}.
The exclusion curve of the KARMEN experiment
lead many people to believe that the LSND evidence
in favor of $\bar\nu_\mu\to\bar\nu_e$ oscillations
was almost ruled out by the result of the KARMEN experiment,
in spite of the fact that the exclusion potential
of the KARMEN experiment was about four times larger than the
actual upper limit.
This discrepancy was due to the observation of less events
than the expected background.
In 1999 the KARMEN experiment
reported the observation of as many events as expected from background
\cite{Jannakos-99},
resulting in an upper limit practically
coincident with the exclusion potential,
compatible with the results of the LSND experiment
(see also \cite{Eitel-99}).
It is clear that the result presented in 1998
has been worse than useless:
its only effect has been to confuse people.
This confusion could have been avoided if the
KARMEN collaboration would not have presented
the 1998 exclusion curve obtained with the Unified Approach,
since it was clearly meaningless from a physical point of view
(although statistically correct if the KARMEN collaboration
did not choose the Unified Approach on the basis of the data,
for example because it gave a bound more stringent than other methods).
Moreover,
the KARMEN collaboration
did not
\cite{KARMEN-nu98}
(and still does not, whereas curiously
they continue to give the useless 1998 exclusion curve \cite{Jannakos-99})
present the result of a Bayesian analysis,
which is less sensitive to fluctuations of the background
than the Unified Approach
(see, for example, \cite{Astone-clw,Giunti-Laveder-00}).
This is probably a negative consequence of the
above mentioned (question~\ref{hot})
bias of the 1998 edition of RPP towards the Unified Approach.

Another lesson to be learned from the KARMEN example
regards the usefulness of the goodness of fit test
proposed by Feldman and Cousins \cite{Feldman-Cousins-98}:
in the Poisson with background case the natural analogue for the
goodness of fit is the probability to obtain $n \leq n_{\mathrm{obs}}$
under the best-fit assumption of $\mu=0$,
where $\mu$ is the mean of signal events,
$n$ is the number of events
and $n_{\mathrm{obs}}$ is the number of observed events.
In the case of the KARMEN experiment
$n_{\mathrm{obs}}$ was zero in 1998
and
the probability to obtain $n = 0$
with $\mu=0$
and a known background of
$2.88 \pm 0.13$
events is 5.6\%.
This probability is not unacceptable,
but imagine that we decide to reject a goodness of fit lower than 10\%.
The problem is that in this case there is nothing to reject,
because the background is known \cite{Armbruster-now98}
and low fluctuations of the background are allowed.
Thus,
in the case of a Poisson process with know background
the goodness of fit test
proposed by Feldman and Cousins
is useless.

\item
\textbf{The confidence level
must be chosen independently from the knowledge of the data?}
\label{confidence level}

As far as I know,
this important question has not been discussed in the literature.
I think that the answer depends on the use
that one is going to make with the confidence intervals.

If many experiments have been done,
the resulting confidence intervals at a certain confidence level
form a set that covers the true value of the quantity under measurement
with probability given by the confidence level.
This property is damaged
if the confidence level is chosen on the basis of the data.
For example,
in the case of a Poisson process with known background,
it is reasonable to choose a higher confidence level
if less events than the expected background have been observed,
because the low fluctuation of the background induces a certain skepticism
on the reliability of the confidence interval.
But in this way
the set of confidence intervals with high confidence level
is unbalanced towards low values of the quantity under measurement,
whereas
the set of confidence intervals with low confidence level
is unbalanced towards high values of the quantity under measurement.
Thus,
the sets of confidence intervals do not have
correct coverage
and the answer to the question above is ``yes''.

In practice, however,
at least in high energy physics research,
one does not have the possibility to do many experiments
for the measurement of a certain quantity
and one is not interested in collecting a set of confidence intervals
that cover the true value of the quantity under measurement
with a given probability.
As discussed above in \ref{usefulness},
each experimental collaboration is interested to
obtain a meaningful and reliable result in its experiment.
Nobody is going to collect sets of confidence intervals
obtained in different experiments and study their properties.
The confidence interval obtained in each experiment is considered individually,
not embedded in a set.
In this case the confidence level
can be chosen after the data are known
without spoiling coverage.

It is now well known that the method to construct the confidence belt
must be chosen independently of the knowledge of the data
\cite{Feldman-Cousins-98}.
A simple reason
is that knowing the data one can
always construct a confidence belt
that gives any wanted confidence interval
at some (sometimes small) confidence level.
But when the method to construct the confidence belt
has been chosen
the coverage of the confidence belt is guaranteed for any
value of the confidence level
and the freedom to choose the confidence level does not allow
one to get a wanted confidence interval.

Taking into account that the confidence interval
can be chosen at will,
I think that it is highly desirable that
the experimental collaborations publish not
a single confidence interval at an arbitrary confidence level,
but the entire \emph{confidence distribution} of the parameter
\cite{Cox-58},
\textit{i.e.} the limits of the confidence interval as functions of
the confidence level,
at least for large values of the confidence level
(say larger than 68\%).
In these days this can be easily done
even for multi-dimensional confidence intervals
by giving a table available as a file through the Internet
and/or
an interpolating function.

\end{enumerate}

In conclusion,
I have discussed some crucial questions regarding
Frequentist confidence intervals.
I hope that the answers that I have given
will at least stimulate the debate on the subject and,
if they are right,
will contribute to an improvement of the understanding
of the usefulness of confidence intervals.

\end{document}